# A Degree Centrality in Multi-layered Social Network

Piotr Bródka[1,2], Krzysztof Skibicki[1], Przemysław Kazienko[1,2], Katarzyna Musiał[3]

[1] Institute of Informatics, Wrocław University of Technology, Wyb.Wyspiańskiego 27, 50-370 Wrocław, Poland
[2] Research Engineering Center Sp. z o.o., ul. Strzegomska 46B, 53-611 Wrocław, Poland
[3] School of Design, Engineering & Computing, Bournemouth University, Poole, Dorset, BH12 5BB, United Kingdom
piotr.brodka@pwr.wroc.pl, krzysztof.skibicki@gmail.com, kazienko@pwr.wroc.pl, kmusial@bournemouth.ac.uk

*Abstract*— **Multi-layered social networks reflect complex relationships existing in modern interconnected IT systems. In such a network each pair of nodes may be linked by many edges that correspond to different communication or collaboration user activities. Multi-layered degree centrality for multi-layered social networks is presented in the paper. Experimental studies were carried out on data collected from the real Web 2.0 site. The multi-layered social network extracted from this data consists of ten distinct layers and the network analysis was performed for different degree centralities measures.**

*Keywords-social network, social network analysis, multi-layered social network, multi-layered neigbourhood, degree centrality, cross-layered degree centrality*

## I. INTRODUCTION

A social network consists of nodes (social entities: humans or groups of people) and relationships (edges) linking pairs of nodes [5], [11]. Nowadays, researchers explore a new social media, in particular web-based services and analyse them using social networks as a model and social network analysis methods as a scientific tool. The analytical thinking about human relationships may involve only one relation type i.e. one kind of connection between users, for example simple friend relationships [1], links based on email exchange [2], [4], or computer networks modelled as simple social networks [12]. However, modern social media allow users to interact and collaborate with each other in many different ways directly or indirectly. For instance, a social networking service enables to publish photos, comment them, mark as favourite, tag them, add other user to contact list, join the user groups, comment the profile or photo, categorize photo, post in topics, etc. [9], [10]. That shows that social systems may be very complex [10]. Due to that fact, new analysis conformed to such complex data need to be developed.

This paper is addressed to special kind of networks called multi-layered social networks. The necessity of a new structure has come out when new web sites (Web 2.0), in which users significantly contribute to their content, became more and more complex and offered to users' variety of interactions and cooperation [10]. The rapid growth of the amount of monitored user activity per minute and aggregated data collected for longer periods (years) as well as the variety of different activity types [7], lead to a point where all activities cannot be treated in the same way. Thus, the different collaboration and communication types should be handled differently, even though they may depend on each other. For example, one user $x$ posts a video in YouTube, the second one $y$ comments it and the third $z$ sends it to yet another user. In this example three different types of activities were enumerated. Therefore, one-layered social network structure is insufficient as multi-layered one emerges as a way of depicting the complexity of relations between users [6]. This paper covers and studies the complex neighbourhoods of nodes within the multi-layered network.

## II. MULTI-LAYERED SOCIAL NETWORK

**Definition**: A multi-layered social network (*MSN*) is defined as a tuple $<V,E,L>$, where: $V$ – is a not-empty set of nodes (social entities: humans, organizations, departments etc.); $E$ – is a set of tuples $<x,y,l>$, $x,y \in V$, $l \in L$, $x \neq y$ and for any two tuples $<x,y,l>$, $<x',y',l'> \in E$ if $x=x'$ and $y=y'$ then $l \neq l'$; $L$ – is a set of distinct layers.

Each tuple $<x,y,l>$ is an edge from $x$ to $y$ in layer $l$ in the multi-layered social network (MSN). The condition $x \neq y$ preserves from loops, i.e. reflexive relations from $x$ to $x$. Moreover, there may exist only one edge $<x,y,l>$ from $x$ to $y$ in a given layer $l$. That means any two nodes $x$ and $y$ may be connected with up to $|L|$ (cardinality of a set $L$) edges coming from different layers. Edges in *MSN* are directed and for that reason, $<x,y,l> \neq <y,x,l>$. Each layer corresponds to one type of relationship between users [9]. Examples of different relationship types can be real world friendship, Facebook friendship, family bonds or work ties. A separate relationship can also be defined based on distinct user activities towards some 'meeting objects', for example publishing photos, commenting photos, adding photos to favourites, etc. (see [9] for details). Depending on variety of user activities types, the *MSN* will consists of more or less layers.

Nodes $V$ and edges $E$ from only one layer $l \in L$ (such edges form set $E_l$) correspond to a simple, one-layered social network $SN <V, E_l, \{l\}>$.

A multi-layered social network $MSN=<V,E,L>$ can be represented by a directed multi-graph. In Figure 1, the example of three-layered social network is presented. The set of nodes consists of $\{x, y, u, v, z, t\}$ so there are five users in the network that can be connected with each other within three layers: $l_1$, $l_2$ and $l_3$. Taking into account layer $l_1$, eight relationships between users: $<x,y,l_1>$, $<y,x,l_1>$, $<x,z,l_1>$,

<$z,x,l_1$>, <$y,z,l_1$>, <$u,z,l_1$>, <$u,v,l_1$>, <$v,u,l_1$> can be distinguished.

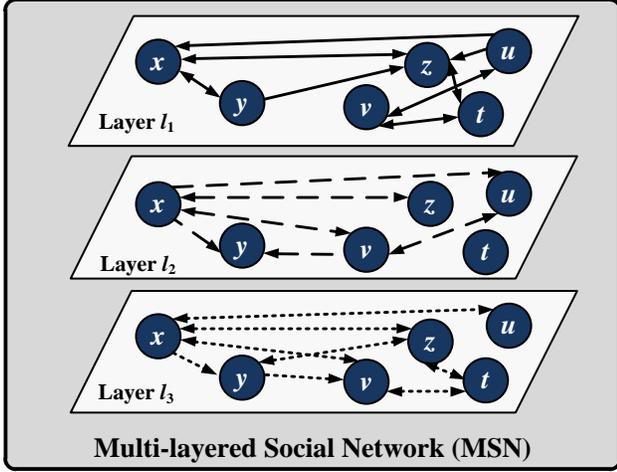

Figure. 1. An example of the multi-layered social network MSN

### III. NEIGHBOURHOOD

Multi-layered social networks are composed with regular social networks (layers) therefore complex neighbourhood set that will span all over the layers has to be defined. However, local neighbourhood can be considered only for one layer. Hence, two different types of neighbourhoods are defined below.

#### A. Local Neighbourhood

Let us consider a neighbourhood in *MSN* but concentrate only on one, particular layer. It is called a local neighbourhood, which is basically equivalent to simple, regular one-layered social network neighbourhood.

$$N(x,l) = \{y : <y,x,l> \in E \vee <x,y,l> \in E\}$$

where: $x$ is a node, $x \in V$, $l$ is a layer, $l \in L$.
For example for *MSN* from Figure 1 the local neighbourhood of node $z$ in each layer is as follows:
- $N(z,l_1) = \{x,y,t,u\}$,
- $N(z,l_2) = \{x\}$,
- $N(z,l_3) = \{x,y,t\}$.

#### B. Multi-layered Neighbourhood

Multi-layered neighbourhood, $MN(x,\alpha)$, of a given node $x$, is a set of nodes that are directly connected with node $x$ on at least $\alpha$ ($1 \le \alpha \ge |L|$) layers in *MSN*:

$$MN(x,\alpha) = \{y : |\{l : <x,y,l> \in E \vee <y,x,l> \in E\}| \ge \alpha\}$$

In the examplary social network from Figure 1, the neighbourhoods of node $x$ for $\alpha$ equals 1, 2, and 3 respectively, are as follows:
- $MN(x,1) = \{u,v,y,z\}$,
- $MN(x,2) = \{u,v,y,z\}$,
- $MN(x,3) = \{u,y,z\}$.

### IV. MULTI-LAYER DEGREE CENTRALITY

Degree centrality indicates relative importance of a node within the network. In general, for a given node $x$ it is calculated as a ratio between number of nodes connected with node $x$ and total number of all nodes in the network (decreased by one). Cross-layer degree centrality (*CLDC*) is defined as a sum of edge weights both incoming to and outgoing from node $x$ towards multi-layered neighbourhood $MN(x,\alpha)$ divided by the number of layers and total number of network members:

$$CLDC(x,\alpha) = \frac{\sum_{y \in MN(x,\alpha)} w(x,y,l) + \sum_{y \in MN(x,\alpha)} w(y,x,l)}{(m-1)|L|}$$

where: $w(x,y,l)$ – the weight of edge <$x,y,l$>.

Similarly to regular degree centrality, we can define cross-layer indegree centrality $CLDC^{In}(x,\alpha)$ in the multi-layered social network *MSN*:

$$CLDC^{In}(x,\alpha) = \frac{\sum_{y \in MN(x,\alpha)} w(y,x,l)}{(m-1)|L|}$$

and cross-layer outdegree centrality $CLDC^{Out}(x,\alpha)$:

$$CLDC^{Out}(x,\alpha) = \frac{\sum_{y \in MN(x,\alpha)} w(x,y,l)}{(m-1)|L|}$$

The value of $CLDC(x,\alpha)$ depends on the parameter $\alpha$, which determines the multi-layered neighbourhood of a given social network member $x$.

### V. EXPERIMENTS

#### A. Real-world Multi-layered Social Network

Data used in the experiments was provided by large polish social network site called extradom.pl. *MSN*s were obtained during transformation from raw data structure to data structure that is presented further. Collected data covers period of 17 months.

TABLE I.    NUMBER OF NODES AND EDGES PER LAYER

| Layer no. | Layer name and roles of users in layer | No. of edges | No. of active nodes |
|---|---|---:|---:|
| 1 | Photo categories | 1,668,060 | 3,826 |
| 2 | Photos comment, author – commenter | 18,130 | 9,606 |
| 3 | Contact list, author – contact | 12,682 | 5,985 |
| 4 | Posts in topic, poster – poster | 1,678,540 | 8,674 |
| 5 | Favourite photos, author-admirer | 182,376 | 12,110 |
| 6 | Photo comments, commenter-commenter | 53,352 | 8,202 |
| 7 | Profile comment, author – commenter | 12,879 | 5,560 |
| 8 | Profile comment, commenter-commenter | 1,156,556 | 3,736 |
| 9 | User in groups, user - user | 7,513 | 11,408 |
| 10 | Contact list, contact – contact | 351,876 | 8,688 |
|  | **SUM** | **5,141,964** | **77,795** |

The extracted multi-layered social has 103,112 nodes and 5,141,964 edges. Distribution of nodes and edges per layer is presented in Table 1.

In the dataset there are 103,112 nodes in total. However there are 77,795 users, who are active, what was revealed after analysing the created layers. It means that 75% of users are active, i.e. have done at least one activity (have a friend in contacts, post, add a picture, etc.). Moreover, only 17,865 users have performed at least one action on one of the layers. Additionally, layers are not equally populated – only three layers have a number of edges greater than 1 million (layer 1, 4, and 8 from Table I). What is interesting, profile comment layer with equal user roles (commentator-commentator) has almost one million edges.

### B. Multi-layered Neighbourhood

Only 17,865 nodes have non-empty set of multi-layered neighbourhood. In Figure 2 distribution of nodes $MN$ for $\alpha=1$ is shown. We can see that only small number of nodes have more than 1000 multi-layered neighbours – it is 5.41%. 16.86% are between 100 and 1000 neighbours, 25.28% has more than 10 and less than 100 adjacent nodes and 52.44% have less than 10 neighbours. Distribution shows typical behaviour of social networks (and $MSN$ as well) which says that the large-scale networks are scale-free ones [3] and their distributions are power law type. It can be seen in Figure 2 that the analysed network features scale-free type distribution.

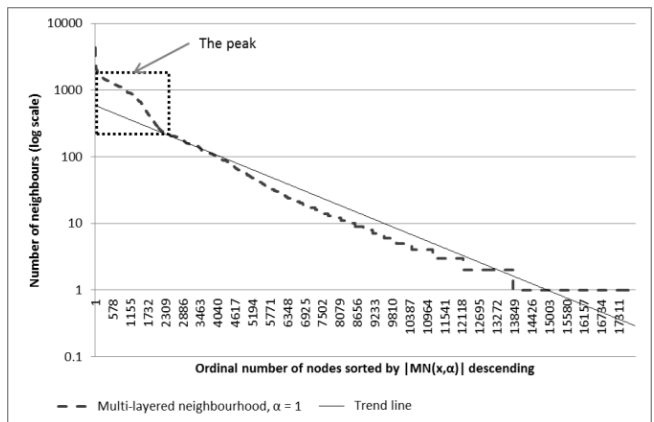

Figure 2 Multi-layer neighbourhood, $\alpha = 1$

The stair-like distribution that appeared after node number 8000 and the amount of neighbour below 10 indicates nodes of the same number of adjacent nodes. To explain that let us look at Figure 5 where distribution of local neighbourhood of all layers is shown. Nodes in layer 2, 3, 7 and 8 have a tendency to have the same number of neighbours. Distributions of multi-layered neighbourhood in layers 4, 5, 6 and 10 are also stair-like distribution however it is not so clearly visible as in layers 2,3,7, and 8. Only layer 9 seems to be great diverse in neighbours number

The peak in the neighbourhood size in the Figure 1 is marked inside the dotted rectangle. It covers around 2,000 nodes. The peak appears due to layer 1 that is shown in Figure 5 – for which the distribution of multi-layered neighbourhood is different than for others. Also layers 9 and 10 contribute to this phenomenon. In each of these three layers almost 2,000 nodes have more than 1,000 neighbours.

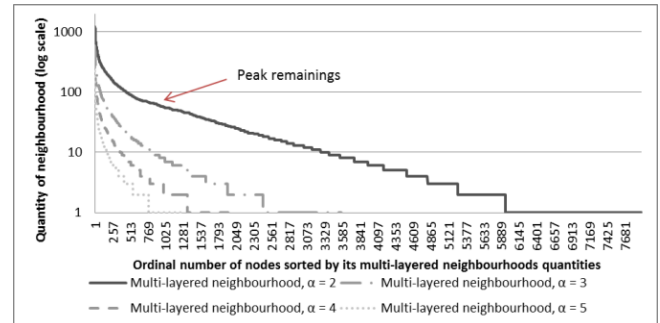

Figure 3 Aggregated log-scaled distribution of Multi-layer neighbourhood, $\alpha$ from 2 to 5

The distributions shown in Figure 3 (for $\alpha=2,3,4,$ and 4) and Figure 4 (for $\alpha=6,7,8,9$ and 10) are similar, all tends to be power-law like, however, the more restricted neighbourhood is (higher $\alpha$) the less power-law it seems to be. It is caused by decreasing number of nodes in those neighbourhoods. On the other hand, the stair-like behaviour starts dominating and is more visible for greate $\alpha$ values. The decreasing number of nodes in $MN$, for greater $\alpha$, is apparent in all distributions.

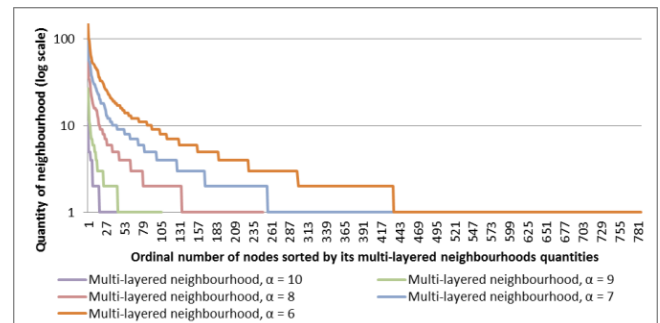

Figure 4 Aggregated log-scaled distribution of Multi-layer neighbourhood, $\alpha$ from 6 to 10

In Figure 5, charts of local neighbourhoods distribution for each layer in $MSN$, are presented. As it is shown great variety appeared among the layers. Power-law distribution is noticeable, however, there is deviation – i.e. layer 1 encounter significant drop down or like those nodes that have less than 10 neighbours are presenting stairs-type distribution.

In Figure 5 we can see that for all layers minimum number of neighbours (above zero) is two, i.e. there is no layer that would have node with one neighbour only.

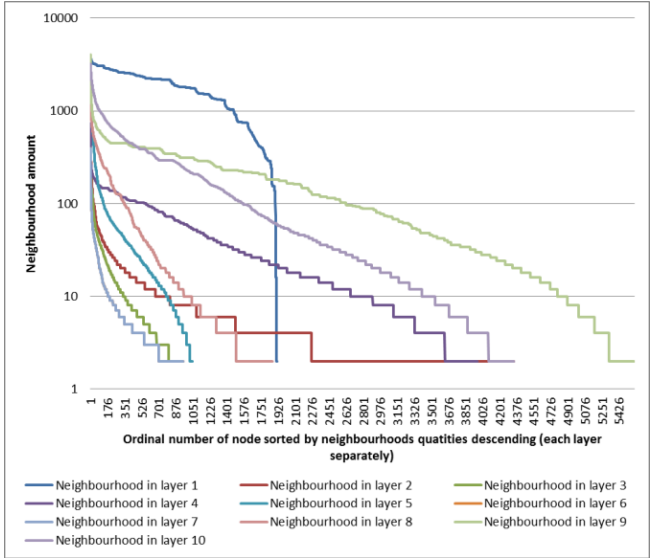

Figure 5 Local neighbourhoods, $LN(x, 1)$

Figure 6, 7 and 8 show histograms where axis $OX$ – ranges – indicates size of multi-layer neighbourhoods and axis $OY$ – quantity – number of nodes with such $|MN(x,\alpha)|$. Shift of first ten ranges is 10 (Figure 6), next ten – 100 (Figure 7), and the last ones –1000 (Figure 8). It let us notice that starting from the beginning of the first range, i.e. from 1 to 10 neighbours, number of nodes are decreasing until it reach the 90 to 100 neighbours. Then it starts increasing rapidly and again going down until it reach 700-800 neighbours.

Another fact to notice is distribution of nodes neighbourhood sizes. First 10 buckets (nodes with less than 100 neighbours, Figure 6) consist of more number of nodes than the next 10 buckets (nodes with more than 100 and less than 1000 adjacency nodes, Figure 7). Analogous situation is for the last 4 buckets. It means that at average, nodes have less than 100 neighbours.

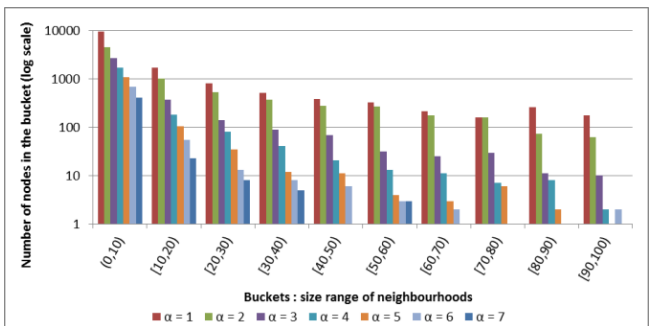

Figure 6 Histogram of Multi-layer neighbourhoods, α from 1 to 7.

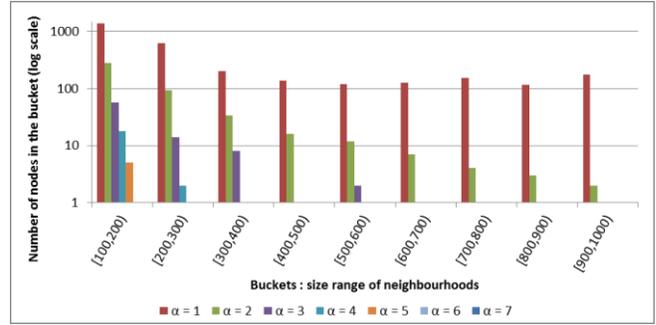

Figure 7 Histogram of Multi-layer neighbourhoods, α from 1 to 7, 2

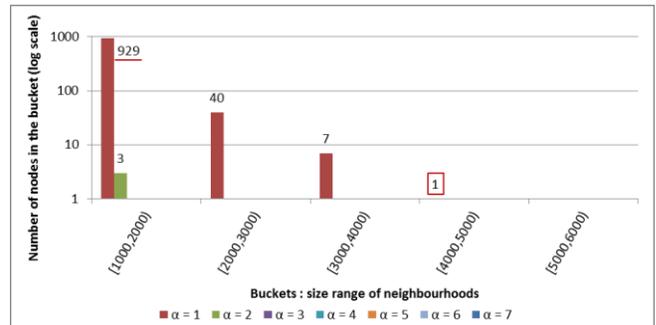

Figure 8 Histogram of Multi-layer neighbourhoods, α from 1 to 7, 3

Last ascertainment is considering only non-empty multi-layered neighbourhoods, which is at most 17% of all nodes in $MSN$1. Figure 9 presents percentage contribution of empty and non-empty sets of $MN(x,\alpha)$.

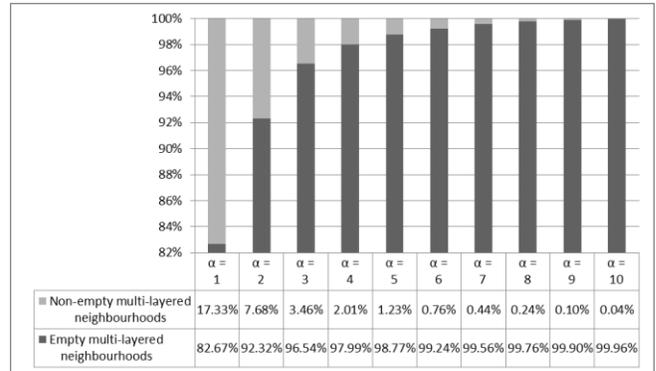

Figure 9 $MSN$1. Percentage of empty and non-empty multi-layer neighbourhoods for $\alpha$ from 1 to 10

The power-law distribution of non-empty multi-layered neighbourhoods (as well as empty ones) is present in Figure 9. It is visible that minor numbers of nodes are interacting with others.

### C. Cross-layer Degree Centrality

Cross-layer degree centrality was calculated separately for each $\alpha \in [1;10]$. Number of nodes that have non zero value of $CLDC(x, \alpha)$ depends on $MN(x, \alpha)$, i.e. there is the same number of nodes having $CLDC(x, \alpha) > 0$ as $|MN(x, \alpha)| > 0$. Additionally, the value of metric is strictly connected with multi-layered neighbourhood.

We can see in Figure 10 the distribution of cross-layer neighbourhood for *α*=1 is asymptotically to power-law, however we can spot the peak (or rather a plateau in lin-log scale; the peak is noticeable in attached to the Figure 10 lin-lin chart) that is marked between vertical lines. To explain this phenomena let us go back to the Figure 2 where the peak was also noticed. Firstly, we said the plateau indicates the temporary stop of a neighbourhood sizes decrease. Secondly, the greater value of the weight means less number of neighbours (the fewer adjacent node the more care is given to them) or greater activity in comparison to other nodes (i.e. commenting the same photo or profile).

The *CLDC*(*x*, 2) shown in Figure 11 is similar to previous one, however for smaller number of nodes. It ranges from 1,07E-04 to 2,43E-10 which is also smaller range than in the case of *CLDC*(*x*, 1). Also the peak is still visible but behaves this time more like peak than plateau (little break of decrease tendency, no linear as it was in the previous figure). However, a phenomenon is not apparent in the attached to Figure 11 lin-lin scale chart.

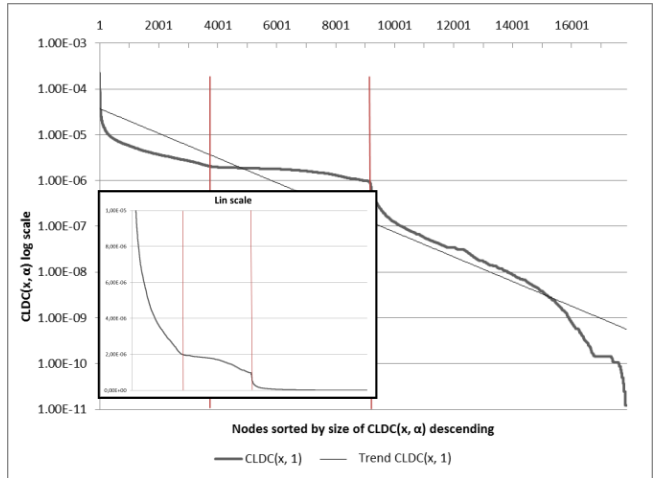

Figure 10 *CLDC*(*x*, 1) distribution

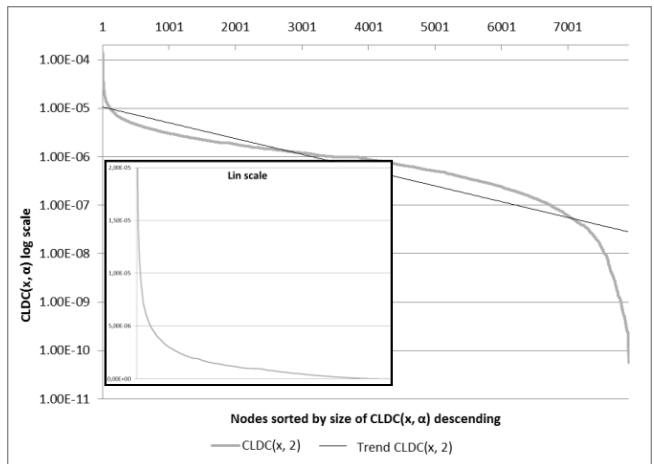

Figure 11 *CLDC*(*x*, 2) distribution

Last distributions for *α*∈[3;10] (Figure 12 and Figure 13) are similar to the ones presented above, i.e. their distributions are asymptotically to power-law. Ranges are shown in Table II.

TABLE II. MINIMUM AND MAXIMUM VALUES OF CLDC(x, α)

| α | 1 | 2 | 3 | 4 | 5 |
|---|---|---|---|---|---|
| Max | 2,25E-04 | 1,47E-04 | 1,07E-04 | 7,66E-05 | 4,87E-05 |
| Min | 1,23E-11 | 2,43E-10 | 2,43E-10 | 1,83E-09 | 7,61E-09 |
| α | 6 | 7 | 8 | 9 | 10 |
| Max | 3,49E-05 | 2,40E-05 | 1,65E-05 | 8,30E-06 | 1,92E-06 |
| Min | 1,39E-08 | 4,52E-08 | 5,43E-08 | 7,40E-08 | 7,40E-08 |

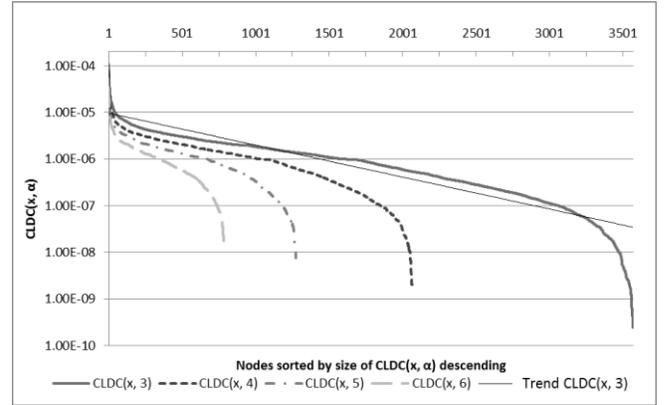

Figure 12 CLDC(x, α) distribution, α = {3, 4, 5, 6}

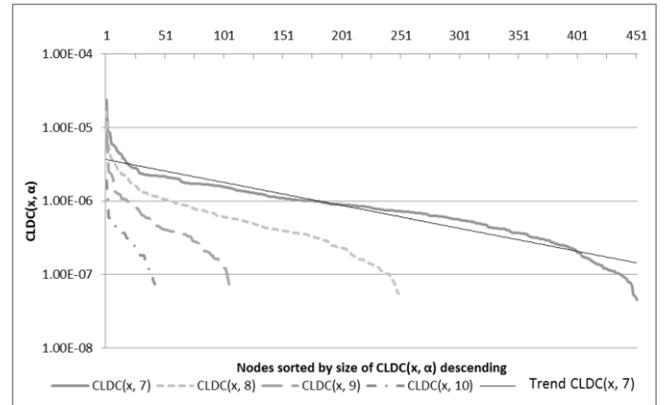

Figure 13 CLDC(x, α) distribution, α = {7, 8, 9, 10}

## VI. CONCLUSIONS AND FUTURE WORK

Three different degree centralities for multi-layered social networks are proposed in the paper: cross-layer degree centrality (*CLDC*), cross-layer indegree centrality (*CLDC$^{In}$*), and cross-layer outdegree centrality (*CLDC$^{Out}$*). They are new structural measures for multi-layered social networks useful in complex social network analysis. Their parameter *α* – the minimum number of layers containing edges – enables the adjustment of the measures to the analyst needs. Obviously, the greater *α* the lower degree centrality values.

The future work will focus on studies using different data sets as well as application of these measures to collective classification problem as label-dependent features [8]. Another research direction is development of efficient algorithms to calculate the measures for huge social networks.

ACKNOWLEDGMENT

The work was supported by Fellowship co-Financed by European Union within European Social Fund and The Polish Ministry of Science and Higher Education the research project, 2010-2013.

The publication has been prepared as part of the project of the City of Wrocław, entitled ― "Green Transfer" -- academia-to-business knowledge transfer project co-financed by the European Union under the European Social Fund, under the Operational Programme Human Capital (OP HC): sub-measure 8.2.1.